\documentclass[12pt]{article}
\usepackage{epsfig,amsfonts,amssymb}
\usepackage{hyperref}
\usepackage{cite}
\input epsf.sty
\usepackage{cite}

\def\ZZZ{{\hbox{ Z\kern-1.6mm Z}}}
\def\RRR{{\hbox{ R\kern-2.4mm R}}}
\def\CCC{{\hbox{ C\kern-2.0mm C}}}
\def\zzz{{\hbox{z\kern-1mm z}}}

\newcommand{\qeq}{{\hbox{=\kern-2.3mm ? \kern.5mm }}}
\renewcommand{\qeq}{=}

\newcommand{\eps}{\epsilon}

\newcommand{\FF}{{\cal F}}

\newcommand{\OO}{{\cal O}}

\newcommand{\EE}{{\cal E}}
\newcommand{\LL}{{\cal L}}

\newcommand{\wt}{\widetilde}
\newcommand{\wh}{\widehat}

\newcommand{\NN}{{\cal N}}

\newcommand{\be}{\begin{equation}}
\newcommand{\ee}{\end{equation}}
\newcommand{\ben}{\begin{eqnarray}\displaystyle}
\newcommand{\een}{\end{eqnarray}}

\newcommand{\refb}[1]{(\ref{#1})}
\newcommand{\p}{\partial}
\newcommand{\sectiono}[1]{\section{#1}\setcounter{equation}{0}}

\def\one{{\hbox{ 1\kern-.8mm l}}}
\def\zero{{\hbox{ 0\kern-1.5mm 0}}}

\begin{document}

\baselineskip 24pt

\begin{center}
{\Large \bf
Quantum Entropy Function from $AdS_2/CFT_1$
Correspondence}

\end{center}

\vskip .6cm
\medskip

\vspace*{4.0ex}

\baselineskip=18pt

\centerline{\large \rm   Ashoke Sen }

\vspace*{4.0ex}

\centerline{\large \it Harish-Chandra Research Institute}

\centerline{\large \it  Chhatnag Road, Jhusi,
Allahabad 211019, INDIA}

\vspace*{1.0ex}
\centerline{E-mail:  sen@mri.ernet.in, ashokesen1999@gmail.com}

\vspace*{5.0ex}

\centerline{\bf Abstract} \bigskip

We review and extend
recent attempts to find a precise relation between
extremal black hole entropy and degeneracy of microstates 
using $AdS_2/CFT_1$ correspondence. 
Our analysis leads to a specific relation between degeneracy
of
black hole microstates and an appropriately defined partition
function of string theory on the near horizon geometry, -- named
the quantum entropy function. In the classical limit this reduces
to the usual relation between statistical entropy and Wald entropy.


\vfill \eject

\baselineskip=18pt

\tableofcontents

\sectiono{Introduction and Motivation} \label{s0}

One of the successes of string theory has been an explanation
of the Bekenstein-Hawking entropy of a class of supersymmetric 
black holes
in terms of microscopic quantum states. In particular Strominger and
Vafa\cite{9601029} 
computed the Bekenstein-Hawking entropy of a class of
extremal supersymmetric black
holes in type IIB string theory on 
$K3\times S^1$ via the formula
\be \label{ei1}
S_{BH} = {A\over 4 G_N}\, ,
\ee
and found agreement with the statistical entropy defined
via the relation
\be \label{ei2}
S_{micro}= \ln d_{micro}\, .
\ee
In \refb{ei1} $A$ is the area of the event horizon, $G_N$ is the
Newton's constant and we have set $\hbar=c=1$. In \refb{ei2}
$d_{micro}$ is the number of microstates of a D-brane system
carrying the same quantum numbers as the black hole. Similar agreement between $S_{BH}$ and $S_{micro}$ has 
been found in a large class of other 
extremal black
holes.
This provides us with
a statistical interpretation of the Bekenstein-Hawking entropy.

The initial studies of $S_{BH}$ and $S_{micro}$ was done in the
limit of large charges. In this limit the calculation simplifies on both
sides. On the black hole side the curvature at the horizon is small, and
we can work with two derivative terms in the full string effective
action. On the microscopic side we can use the asymptotic formula for
$d_{micro}$ for large charges instead of having to compute it exactly.
However it is clearly of interest to know if the correspondence between
the black hole entropy and the statistical entropy extends beyond the
large charge limit. 

In order to address this problem we need to open two fronts. First of
all we need to compute the degeneracy of states of  black
holes to greater accuracy so that we can compute corrections
to $S_{micro}$ given in \refb{ei2}. Conceptually this is a
straightforward problem since
$d_{micro}$ is a
well defined number, 
especially in the case of BPS extremal black holes, since the
BPS property gives a clean separation between the spectrum of
BPS and non-BPS states.  Technically, counting of $d_{micro}$
is a challenging problem, although 
this has now
been achieved for a class of black holes 
in $\NN=4$ supersymmetric string
theories\cite{9607026,0510147,0609109,0802.1556}. 
Significant 
progress has also been made for half BPS black holes in
a class of
$\NN=2$ supersymmetric string 
theories\cite{0602046,0702146,0711.1971}. 

The other front involves 
understanding
how higher derivative corrections / string loop corrections affect the
black hole entropy. 
Wald's formalism\cite{9307038,9312023,9403028,9502009}
gives a clear prescription for calculating the effect of tree level 
higher derivative corrections on the black hole entropy,
and for extremal black holes this leads to the
entropy function formalism\cite{0506177,0606244,0708.1270}. 
Thus here there is no conceptual problem, but in order to implement
it we need to know the higher derivative terms in the 
action.\footnote{Some of these corrections, which can be regarded
as the correction to the central charge in the Cardy formula, have now
been computed on both sides and the results 
match\cite{9711053,9801081,9812082,0007195,0409148,0506176}.}
Inclusion of quantum corrections into the computation of 
black hole
entropy
is more challenging both
conceptually and technically, 
and
this will be the main issue we shall try to address.

We  restrict our study to extremal black
holes with a near horizon geometry of the form
$AdS_2\times K$, where $K$ is a compact space. We shall
argue that the key to defining the quantum corrected black hole
entropy is in the partition function $Z_{AdS_2}$ of string theory on 
$AdS_2\times K$\cite{0805.0095,0806.0053}. 
This partition function is divergent due
to the infinite volume of $AdS_2$, but we show 
in \S\ref{s2} that there is an 
unambiguous procedure for extracting its finite part.
In particular we argue that the partition function
has the form $e^{C L}\times$ a finite part where $C$ is a
constant and $L$ is the length of the boundary of regulated
$AdS_2$. We define the finite part as the one obtained by
dropping the $e^{CL}$ part and also argue that this is a natural
prescription from the point of view of the dual 
$CFT_1$.\footnote{An
earlier attempt to extract this finite part can be found in
\cite{0608021}. 
This procedure, which relies heavily on supersymmetry,
is quite different from ours.}
We then argue that $AdS_2/CFT_1$ 
correspondence\cite{9711200,9802109,9802150,9809027}
leads to the following relation between the $AdS_2$ 
partition function and the
microscopic degeneracy $d_{micro}(\vec q)$ of black holes carrying
charge $\vec q\equiv \{q_i\}$: 
\be \label{ealt}
d_{micro}(\vec q) = \left\langle \exp[-
i  q_i\ointop d\theta \, A^{(i)}_\theta]
\right\rangle^{finite}_{AdS_2}\, ,
\ee
where $\langle ~\rangle_{AdS_2}$ 
denotes the unnormalized path integral
over various fields on euclidean global
$AdS_2$ associated with the attractor geometry for charge
$\vec q$ and $A^{(i)}_\theta$
denotes the component of the $i$-th gauge field along the boundary
of $AdS_2$.
This equation gives a precise relation
between the microscopic degeneracy and an
appropriate partition function in
the near horizon geometry of the black hole.
We shall call the right hand side of \refb{ealt} the `quantum entropy
function'. 
In the classical limit this reduces to 
the exponential of the Wald entropy of the black hole.  Thus
\refb{ealt} is the quantum generalization of the 
$S_{BH}=S_{micro}$ relation in the classical theory.
This is the key result of our analysis.

In defining the functional integral over various fields involved
in
computing
the right hand side of \refb{ealt} we need to use 
appropriate boundary condition
on the various fields. For each $U(1)$ gauge field the solution to
the classical equations of motion  locally
has two independent solutions near the boundary; the 
constant mode and the mode representing the asymptotic value
of the electric field (and hence the charge). In $AdS_2$ the
electric field mode is the dominant one near the boundary and hence
we hold this  fixed as we carry out the functional integral over
various fields in $AdS_2$. 
In the classical limit the constant mode is determined in terms
of the asymptotic electric field by requiring absence of
singularity in the interior of $AdS_2$ whereas in the quantum
theory we integrate over this mode.
Since the asymptotic electric fields determine the charge carried by
the black hole, this boundary condition
forces us to work in a fixed
charge sector, and as a result the dual $CFT_1$ also has states of
a fixed charge. In special circumstances however it may be possible
to define the partition function with fixed values of the constant
modes of the gauge field at the boundary. Classical equations of
motion together with the requirement of absence of singularities in
the interior of $AdS_2$ 
sets the asymptotic electric field to be equal to the
constant mode; so we shall continue to denote by $e_i$ the boundary
values of the constant modes. 
Quantum mechanically of course we cannot fix both modes; so we have
to allow the electric field modes to fluctuate.
This leads to a new  
partition function
$Z_{AdS_2}(\vec e)$. If we denote by 
$Z^{finite}_{AdS_2}(\vec e)$ the finite part of this partition
function then $AdS_2/CFT_1$ correspondence suggests the
following relation between this partition function and the
microscopic degeneracies of the boundary theory:
\be \label{egen1}
Z^{finite}_{AdS_2}(\vec e) 
= \sum_{\vec q} d_{micro}(\vec q) \,
e^{-2\pi \vec e\cdot \vec q} \, .
\ee
Note that the right hand side now has a sum over different charges
since in the definition of the $AdS_2$ partition function we have 
allowed the
asymptotic electric fields to fluctuate.
We show that
\begin{enumerate}
\item In the classical limit eq.\refb{egen1} again
gives us back the
relation equating the classical Wald entropy   
and statistical entropy.
\item For black holes in type IIA string theory compactified on a
Calabi-Yau 3-fold, eq.\refb{egen1} is closely related to the
OSV conjecture\cite{0405146}.
\end{enumerate}
However
$Z^{finite}_{AdS_2}$
appearing in \refb{egen1} may be defined only under special
circumstances since here we fix the constant modes of the 
asymptotic gauge fields and allow the non-normalizable modes
corresponding to the asymptotic electric fields
to fluctuate. Indeed, in \S\ref{s6} 
we point out various subtleties with this
formula that makes it clear that even when such a partition
function can be defined, it probably only makes sense as an
asymptotic expansion around the classical limit.

Since the classical counterpart of our results
is encoded in the
entropy function formalism for computing entropy of extremal
black holes\cite{0506177,0606244,0708.1270}, 
we begin with a lightening review of 
this formalism in \S\ref{s1}. 
This takes into account
the effect of tree level
higher derivative corrections on the computation of black hole
entropy.

\sectiono{The Extremal Limit and the Entropy
Function Formalism} \label{s1}

Let us consider the metric of a Reissner-Nordstrom black hole
in (3+1) dimensions. It is given by
\be \label{em1}
ds^2  = - (1 - a/\rho) (1 - b/\rho) d\tau^2   + {d\rho^2\over
 (1 -a/\rho) (1 - b/\rho)}   
 + \rho^2 (d\theta^2 + \sin^2\theta d\phi^2)\, . \ee
Here $(\tau,\rho,\theta,\phi)$ are the coordinates of space-time and
$a$ and $b$ are two parameters labelling the positions
of the outer and inner horizon of the black hole respectively ($a>b$).
The extremal limit corresponds to $b\to a$. 
We take this limit keeping the
coordinates $\theta$, $\phi$, and
  \be \label{em2}
r \equiv 2\left( \rho- {a+b\over 2}\right)/ (a-b), \quad
t\equiv (a-b) \tau / 2a^2 \, , 
\ee
fixed\cite{9812073,9904143}. 
In this limit the metric takes the form:
\be \label{et1}
ds^2 = a^2\left(-(r^2-1) dt^2 + {dr^2\over r^2-1}\right)
+ a^2 (d\theta^2 + \sin^2\theta d\phi^2)\, . \ee
This is the metric of $AdS_2\times S^2$, with
$AdS_2$ parametrized by $(r,t)$ and $S^2$ parametrized by
$(\theta,\phi)$.
Although in the original coordinate system the
horizons coincide in the extremal limit, in the
$(r,t)$ coordinate system the two horizons are
at $r=\pm 1$.

All known extremal black hole solutions have an $AdS_2$ factor
in their near horizon geometry; furthermore the other near horizon
field configurations remain invariant under the $SO(2,1)$ isometry
of $AdS_2$. We shall take this to be the definition of an
extremal black hole even in theories with higher derivative terms
in the action\cite{0506177,0606244,0705.4214,0803.2998}.
We can give a uniform treatment of all such extremal
black holes by regarding the angular directions as part of compact
coordinates. Thus we have an effective two dimensional theory
of gravity coupled to (infinite number of) other fields. 
Among them of
particular importance are the $U(1)$ gauge fields which we shall denote
by $A_\mu^{(i)}$. The most general near horizon geometry
consistent with the $SO(2,1)$ isometries of $AdS_2$ is given by
\be \label{et2}
ds^2
= v\left(-(r^2-1) dt^2+{dr^2\over
r^2-1}\right),  \quad F^{(i)}_{rt} = e_i, \quad \cdots
\ee
where $F^{(i)}_{\mu\nu} = \p_\mu A^{(i)}_\nu - 
\p_\nu A^{(i)}_\mu$ are the gauge field strengths, $v$ and $e_i$ are
constants and $\cdots$ denotes near horizon values of other fields.
The important point is that the $SO(2,1)$ isometry of $AdS_2$ allows
us to write down the most general near horizon field configuration
in terms of some constants $v$, $e_i$ etc.
Note that we have not explicily displayed the magnetic fields; they
represent flux through the sphere labelled by the angular coordinates and
are regarded as compactification data in our current viewpoint.

Let us denote by $\LL^{(2)}$ the two dimensional Lagrangian density
of this classical string theory including all higher derivative corrections.
Since $\LL^{(2)}$ is a scalar, when evaluated on the near 
horizon geometry \refb{et2} it will give a function of the parmeters
$v$, $e_i$ etc. without any dependence on the coordinates $r$ and $t$. 
We now introduce an extra set of variables
$q_i$ in one to one correspondence to the gauge fields
and define a function 
$\EE$ of $\vec q$ and the near horizon parameters  as
\be \label{et3}
\EE(\vec q, v,\vec e, \cdots) \equiv 2\pi \left( e_i \, q_i - v \, \LL^{(2)}
\right)
\ee
where $\LL^{(2)}$ on the right hand side is evaluated in the near
horizon geometry.
Using classical equations of
motion and Wald's modified formula for black hole 
entropy\cite{9307038,9312023,9403028,9502009}
in the
presence of higher derivative terms one can show that
\begin{enumerate}
\item For a black hole with electric charges $\vec q$, all the near
horizon parameters are determined by extremizing $\EE$ with respect
to the near horizon parameters:
\be \label{et4}
{\p \EE\over \p v}=0, \quad {\p \EE\over \p e_i}=0, \quad \cdots\, .
\ee
Since the dependence of $\EE$ on all the near horizon parameters
other than $\vec e$ come only through $v\LL^{(2)}$, we can
first extremize $v\LL^{(2)}$ with respect to these parameters for
a fixed $\vec e$ and call the result $v\LL^{(2)}(\vec e)$. Then we can
extremize $\EE=2\pi (\vec e\cdot \vec q - v\LL^{(2)}(\vec e))$ 
with respect to $e_i$ to find the relation between $\vec e$ and $\vec q$.

\item The Wald entropy is given by the value of $\EE$ at this
extremum:
\be \label{et5}
S_{BH}(\vec q) = \EE\, .
\ee

\end{enumerate}

\sectiono{$AdS_2$ Partition Function} \label{s2}

A solution to eqs.\refb{et4} describes an $AdS_2$ background
via \refb{et2}. 
We can analytically continue the solution to euclidean space
by defining
new coordinates $\theta$ and $\eta$ via
\be \label{et6}
t = -i\theta, \qquad r = \cosh\eta\, .
\ee
In these coordinates the solution \refb{et2} 
takes the form
\be \label{et7}
{ds^2} = { v \, \left(d\eta^2 +\sinh^2\eta \, d\theta^2 \right),}
\qquad
F^{(i)}_{\theta\eta} = i e_i \, \sinh\eta, \qquad \cdots \, .
\nonumber \ee
The metric is non-singular at the
point $\eta=0$ if we choose $\theta$ to have period $2\pi$.
Integrating the field strength we can get the form of the gauge
field:
\be \label{et8}
A_\mu^{(i)} dx^\mu = -i \, e_i \, (\cosh\eta  
{ -1}) d\theta=  -i \, e_i \, (r  
{ -1})d\theta\, .
\ee
Note that the $-1$ factor inside the parenthesis is required to make the
gauge fields non-singular at $\eta=0$.

We can formally define the partition function of string theory in this
background as the result of path integral over all the string fields.
However due to infinite volume of $AdS_2$ this partition function is
{\it a priori} divergent. We shall now describe an unambiguous
procedure for extracting the finite part of this partition function.
We begin by describing this
procedure in the classical limit where the
path integral over the string fields is saturated by the saddle point
corresponding to the classical geometry \refb{et7}.
Let $A$ denote the classical
Euclidean action
\be \label{et10} 
A = A_{bulk} + A_{boundary}
\ee
where
\be \label{et10.5}
A_{bulk} = -\int dr d\theta \, \sqrt{\det g} \, \LL^{(2)} \, ,
\ee
evaluated in the $AdS_2$ background, and
$A_{boundary}$ denotes the possible boundary contribution
to the action. 
Then the classical
Euclidean partition function is given by
\be \label{et11}
Z_{AdS_2} = e^{-A}\, .
\ee
Since $AdS_2$ has infinite volume we need to define \refb{et10.5}
via suitable regularization. For this we introduce a cut-off at 
$\eta=\eta_0$ or equivalently $r=r_0=\cosh\eta_0$. Then the
regularized volume of $AdS_2$ is given by
\be \label{et12}
V_{AdS_2} \equiv \int_1^{r_0} dr \int_0^{2\pi} d\theta\,
\sqrt{\det g} =  2\pi\, v \, (r_0 
{-1}) \, .
\ee
This gives\
\be \label{et13}
A_{bulk} = - (r_0{  -1}) \, 2\pi v \, \LL^{(2)}\, .
\ee
On the other hand $A_{boundary}$ can be estimated by
making a change of
coordinates:
\be \label{et14}
w \equiv r_0 \theta\, , \qquad \xi=\eta_0-\eta\, .
\ee
$w$ has period $2\pi r_0$.
In this coordinate system the field configuration near the
boundary $r=r_0$ take the form:
\ben \label{et15}
{ds^2} &{=}& v \, (d\xi^2 + e^{-2\xi}\, dw^2) 
+ \OO(r_0^{-2}), \nonumber \\
A^{(i)}_w &=& -i \, e_i \, \left(e^{-\xi} { -r_0^{-1}}\right)
+ \OO(r_0^{-2}), \qquad
F^{(i)}_{\xi w} = i e_i e^{-\xi}+ \OO(r_0^{-2})\, . 
\een
Now $A_{boundary}$ can be represented as an intergral of a
local expression constructed from the metric, the gauge field
strength $F^{(i)}_{\mu\nu}$ and their derivatives, integrated
over the boundary coordinate $w$. Since the field configuration is
independent of $w$, the integration over $w$ produces an overall
multiplicative factor of $2\pi r_0$. On the other hand since the metric
and $F^{(i)}_{\mu\nu}$ (and all other fields) are independent of
$r_0$ except for corrections of order $r_0^{-2}$, the integrand
appearing in $A_{boundary}$ will share the same property. Thus
after integration over $w$  we shall get a term proportional to
$r_0$ plus corrections of order $r_0^{-1}$:
\be \label{et16}
A_{boundary} = -K r_0 + \OO(r_0^{-1})\, ,
\ee
for some constant $K$. Substituting \refb{et13} and \refb{et16}
into \refb{et11} we get
\be \label{et17}
Z_{AdS_2} = e^{r_0(2\pi v \, \LL^{(2)}+K) 
{ -2\pi v \, \LL^{(2)}}
 + \OO(r_0^{-1})}\, .
 \ee
 The term linear in $r_0$ in the exponent is ambiguous since it can
 be changed by changing the boundary terms. However the finite
 part in the exponent is independent of the boundary terms and is
 unambiguous. We shall define
 \be \label{et18}
 Z_{AdS_2}^{finite} = e^{-2\pi v \, \LL^{(2)}}\, .
 \ee

Our analysis resembles the euclidean action formalism for describing
black hole thermodynamics\cite{gibhaw,0506176,0604070}. This
formalism was originally developed for non-extremal black holes,
but it has been applied to study extremal black holes in 
\cite{0704.0955,0704.1405}. However in these studies one uses
the Euclidean action of the full black hole solution and not just its
near horizon geometry. Also one cannot directly apply the Euclidean
action formalism on an extremal black hole; one needs to begin with
the results for a non-extremal black hole and then take the 
extremal limit at the end.
In contrast our analysis has been based purely on the
near horizon $AdS_2$ geometry. As we shall see in \S\ref{s3},
this will allow us to interpret $Z_{AdS_2}$
as the partition function of a dual $CFT_1$ living on 
the boundary $r=r_0$.

In quantum theory one 
defines $Z_{AdS_2}$ as the path 
integral over the string fields
in $AdS_2$ 
weighted by $e^{-A}$\cite{9802109,9802150}. 
We can hope to represent the effect of this path integral
by a modification of the lagrangian 
density $\LL^{(2)}$
to an appropriate `effective Lagrangian density'.
In flat space-time the one particle irreducible
action is non-local and hence causes an obstruction to expressing the
action as an integral over a local Lagrangian density. However the
situation in $AdS_2$ background is better since the curvature of
$AdS_2$ produces a natural infrared cut-off. Thus the quantum 
$Z_{AdS_2}$ is expected to be given by an expression
similar to \refb{et17}, with $\LL^{(2)}$ replaced by 
$\LL^{(2)}_{eff}$, and $K$ replaced by some other
constant $K'$. Thus we have
\be \label{et19}
Z_{AdS_2} = e^{r_0(2\pi v \, 
\LL^{(2)}_{eff}+K') +  \OO(r_0^{-1})} \, 
Z_{AdS_2}^{finite}\, ,
\ee
where
\be \label{et20}
 Z_{AdS_2}^{finite} = e^{-2\pi v \, \LL^{(2)}_{eff}
 }\, .
 \ee

In the above discussion we have glossed over an important issue.
In order to properly define the path integral over fields
on $AdS_2$ we need to fix the boundary condition on various fields
at $r=r_0$.
 In $AdS_{d+1}$ for general $d$ the classical
Maxwell equations for a gauge
field near the boundary has two independent solutions.
One of these modes represent the constant
asymptotic value of the
gauge fields, the other one measures the asymptotic electric field or
equivalently the charge carried by the
solution. 
Requiring the absence of singularity in the interior of $AdS_{d+1}$
gives a relation between the two coefficients\cite{9802109,9802150}. 
Thus in defining the
path integral over $AdS_{d+1}$ we fix one of the coefficients and
allow the other one to fluctuate.  For $d\ge 3$ the constant mode of the
gauge field is dominant near the boundary; hence it is natural to fix
this and allow the mode measuring the
charge to be determined dynamically in the classical
limit and to fluctuate in the full quantum theory. 
However for $d=1$ the mode
that measures charge is the dominant one near the boundary; thus
it is more natural to think of this as a parameter of the boundary CFT
and let the
constant mode of the gauge field be determined dynamically.
This can be seen for example in \refb{et8} where the term proportional
to $r$ measures the charge and the constant term in the expression for
the gauge field is determined in terms of the linear term by requiring
the gauge fields to be non-singular at the origin. Thus a more natural
definition of the partition function of $AdS_2$ will be to fix the
coefficient of the linear term in $r$ and allow the constant term to
fluctuate.

We shall reserve the symbol $Z_{AdS_2}$ for the  partition 
function defined with fixed value of the constant mode of the
asymptotic gauge field, and $\LL^{(2)}_{eff}$ and $K'$ for the
corresponding quantities which appear in the expression for
$Z_{AdS_2}$ via \refb{et19}. 
Such a partition function may exist only in 
special circumstances since here
we let a non-normalizable mode corresponding to the asymptotic
electric field to fluctuate.
For the more sensible boundary condition where we
fix the asymptotic electric field configuration it will be convenient
to introduce the quantity
\be \label{ehatted}
\wh Z_{AdS_2} 
\equiv
\left\langle \exp[-i  q_i\ointop d\theta \, A^{(i)}_\theta]
\right\rangle_{AdS_2}\, ,
\ee
where $\langle ~\rangle_{AdS_2}$ 
denotes the {\bf unnormalized} path integral
over various fields on $AdS_2$ with fixed
asymptotic values of the electric fields 
corresponding to the attractor
geometry. In the classical theory this is given by 
multiplying the $AdS_2$ partition function 
\refb{et17} and 
$\exp[-i  q_i\ointop d\theta \, A^{(i)}_\theta]$ with 
$A^{(i)}_\theta$
given by its classical value \refb{et8}.  This gives
\be \label{ehat2}
\wh Z_{AdS_2} = \exp\left[ r_0(2\pi v \LL^{(2)}+ K - 2\pi \vec e
\cdot \vec q)
+2\pi (\vec e\cdot \vec q - v \LL^{(2)}) + \OO(r_0^{-1})\right]\, .
\ee
In the quantum theory 
arguments similar to the ones given for $Z_{AdS_2}$ show that
$\wh Z_{AdS_2}$ should have the form
\be \label{ehat3}
\wh Z_{AdS_2} = \exp\left[ 2\, \pi\, C\, r_0
+ \OO(r_0^{-1})\right]
\, \left\langle \exp[-i  q_i\ointop d\theta \, A^{(i)}_\theta]
\right\rangle_{AdS_2}^{finite}\,  ,
\ee
where $C$ is a constant and 
$\left\langle \exp[-i  q_i\ointop d\theta \, A^{(i)}_\theta]
\right\rangle_{AdS_2}^{finite}$
is an $r_0$ independent term. We shall call the latter the
`quantum entropy function'.

\sectiono{$AdS_2/CFT_1$ Correspondence: The
Macroscopic View} \label{s3}

According to \cite{9711200,9802109,9802150} 
string theory on $AdS_{d+1}$ times a compact
space is dual to a $d$ dimensional conformal field theory (CFT)
living on
the boundary of $AdS_{d+1}$. In Euclidean global
$AdS_{d+1}$, the boundary S-matrix of
string theory can be used to construct the correlation functions
of various operators in the $d$-dimensional CFT living on the
boundary $S^d$. This provides
a constructive definition of the $d$-dimensional CFT. 
We shall use the same
procedure to define a one dimensional CFT living on the boundary
of $AdS_2$.
In this case the partition function of string theory
on $AdS_2$ will be identified as the
partition function of the CFT living on the boundary $r=r_0$.

We shall use $w$ defined in \refb{et14}
as the coordinate on the boundary. In this coordinate
system the boundary -- which is a circle -- has period $2\pi r_0$.
Thus as $r_0\to\infty$ the size of the boundary becomes infinite.
On the other hand the metric \refb{et15} near the boundary remains finite
in this limit without acquiring an infinite conformal factor; showing that
the CFT has a finite ultraviolet cut-off. This is somewhat different from
the usual convention where we would use $\theta$ as the coordinate
on the boundary so that the boundary will have fixed
period $2\pi$, and the
$d\theta^2$ term in the metric will have a large conformal factor
proportional to $r_0^2$, indicating that $r_0^{-1}$ provides an
ultraviolet length cut-off in the CFT. Since the partition
function depends on the ratio of the size of the 
boundary and the ultraviolet
length cut-off, it is independent of the convention.

First we consider the case where
we fix the asymptotic value of the electric fields 
$\p_r A^{(i)}_\theta$ to $-ie_i$ as in \refb{et8}. 
This fixes the charge of the black hole;
thus
the dual $CFT_1$ should only contain 
states carrying a fixed charge $\vec q$. 
$AdS_2/CFT_1$ correspondence will relate the partition function
on $AdS_2$ to $Tr\exp(-2\pi r_0 H_w)$ in the dual $CFT_1$ where
$H_w$ denotes the $w$ translation generator. Since we are interested
in taking the $r_0\to\infty$ limit at the end we need to keep in $H_w$
terms up to order $r_0^{-1}$.
Now normally in the presence of a non-zero asymptotic
gauge field, $H_w$ will contain a term 
$iQ_i A_w^{(i)}$ where $Q_i$ is the charge operator conjugate
to the gauge field $A_w^{(i)}$. However here the constant mode
of the $A_\theta^{(i)}$ -- which generates a term
proportional to $r_0^{-1}$ in $A^{(i)}_w$, 
is allowed to fluctuate and does not describe
a fixed background. Hence including this term in $H_w$ 
is not sensible. 
We can instead consider the modified partition function
$\wh Z_{AdS_2}$ introduced in \refb{ehatted}.
Its effect will be to remove the $iQ_i A_w^{(i)}$ term from the
$CFT_1$ Hamiltonian.\footnote{The need to consider 
$\wh Z_{AdS_2}$ instead of $Z_{AdS_2}$ can also be seen
directly on the $AdS_2$ side. In the classical limit
the extremization of the action gives the equations of motion if
we keep fixed the gauge fields $A^{(i)}_\theta$ at the boundary.
However since here we allow the constant mode of the gauge field
to vary, we need to add a boundary term to cancel the term
proportional to $\delta A^{(i)}_\theta$ at the boundary. The
$-iq_i\ointop d\theta\, A^{(i)}_\theta$ term precisely achieves
this purpose.}
We now notice that 
in the $w$ coordinate system the field configuration
\refb{et15} near the boundary is independent of $r_0$ up to corrections
of order $r_0^{-2}$, except for the $ie_i / r_0$ term in $A_w^{(i)}$.
Since the Hamiltonian of $CFT_1$ is determined by the field
configuration on $AdS_2$ near the boundary and since
the Hamiltonian no longer contains any term proportional to
the constant mode of the
gauge field, it must have the form $H + \OO(r_0^{-2})$ where
$H$ is $r_0$ independent. This allows us to write
\be \label{ewrite}
\wh Z_{AdS_2} = Tr \left[\exp\left\{-2\pi r_0 \left(H 
+ \OO(r_0^{-2})\right)\right\}\right]\, .
\ee
For large $r_0$ the right hand side of \refb{ewrite}
should reduce to
\be \label{ek1}
e^{-2\pi E_0 r_0} \, d(\vec q) \, ,
\ee
where $E_0$ is ground state energy and $d(\vec q)$ is the number of
ground states. Comparing this with \refb{ehat3}  we get
\be \label{et23a}
E_0 = -C\, ,
\ee
and
\be \label{ettex}
d(\vec q) = \left\langle \exp[-i  q_i\ointop d\theta \, A^{(i)}_\theta]
\right\rangle^{finite}_{AdS_2}\, .
\ee

Note that in writing down \refb{ek1}
we have implicitly assumed that
$H$ has a discrete spectrum. We can try to relax this assumption by
assuming that $H$ has a continuous spectrum, and that in the sector
with charge $\vec q$ the density of states is
given by some function $f(E,\vec q)$.
This replaces \refb{ek1} by
\be \label{et34}
\int \, dE \, f(E,\vec q) \, e^{-2\pi r_0 E}\, .
\ee
Since this must be equal to $\wh Z_{AdS_2}$ given in
\refb{ehat3} we can use this to determine the form of
$f(E,\vec q)$. Suppose $E_0$ is the ground state energy so that
$f(E,\vec q)$ vanishes for $E<E_0$. 
This gives
\be \label{et35}
e^{-2\pi r_0 E_0} \, 
\int_{E_0}^\infty 
\, dE \, f(E,\vec q) \, e^{-2\pi r_0 (E-E_0) }
= e^{2\, \pi \, r_0\, C +\OO(r_0^{-1})} 
\left\langle \exp[-i  q_i\ointop d\theta \, A^{(i)}_\theta]
\right\rangle^{finite}_{AdS_2} \, .
\ee
Now
$f(E,\vec q)$ depends on $E$ through the
ratio of $E$ to the ultraviolet length cut-off which has been taken to be
of order 1 in our convention. $f$ cannot depend on $r_0$ since by
definition $H$ is $r_0$ independent. 
Comparing the two sides of \refb{et35}
we get
\be \label{eex1}
E_0 = -C\, ,
\ee
and
\be \label{eex2}
\int_{E_0}^\infty 
\, dE \, f(E,\vec q) \, e^{-2\pi r_0 (E-E_0)  }
= \left\langle \exp[-i  q_i\ointop d\theta \, A^{(i)}_\theta]
\right\rangle^{finite}_{AdS_2}\, .
\ee
Since the large $r_0$ behaviour of the integral on the left
hand side will be controlled
by the behaviour of the integrand near $E=E_0$, let us
examine the possible behaviour of the integrand 
near $E=E_0$. 
Since the right hand side of \refb{eex2}
is finite, the integral also
must be finite and hence $f(E,\vec q)$ cannot grow as 
$(E-E_0)^{-\alpha}$ with $\alpha\ge 1$. On the other hand if
it grows as $(E-E_0)^{-\alpha}$ with $\alpha< 1$ the result
of the integral on the left hand side of \refb{eex2}
will behave as $r_0^{\alpha-1}$ for large $r_0$. This vanishes
as $r_0\to\infty$
in contradiction to what we have on the right hand side of this
equation. A simple resolution to this is that $f(E,\vec q)$ has a term
\be \label{et36}
\delta(E-E_0) \, d(\vec q), \qquad   d(\vec q)
  = \left\langle \exp[-i  q_i\ointop d\theta \, A^{(i)}_\theta]
\right\rangle^{finite}_{AdS_2}\, ,
\ee
together with possible additive terms of the form 
$(E-E_0)^{-\alpha}$
with $\alpha<1$ which do not contribute in
the large $r_0$ limit. \refb{et36} implies a discrete set of states
at $E=E_0$ besides a possible continuum. There can also be other
terms proportional to $\delta(E-E_i)$ with $E_i>E_0$; the
contribution from these terms to \refb{eex2} will be
exponentially suppressed. This gives us back \refb{ettex}.

It is instructive to see what the above result for the spectrum 
implies
if we change our viewpoint to the more conventional
one, and interprete the $r_0\to\infty$ limit
as taking the ultraviolet length cut-off to zero at fixed size of the
boundary. If $\eps$ denotes the energy and $\wt f(\eps,\vec q)$ 
denotes the density of states in this new unit, and if we choose the
zero of $\eps$ to be at the ground state, then we have
\be \label{et37}
\eps = (E-E_0) r_0, \qquad \wt f(\eps,\vec q) = r_0^{-1} f(E,\vec q) \, .
\ee
This gives
\be \label{et38}
\wt f(\eps,\vec q) = r_0^{-1} \delta (E-E_0) \, d(\vec q)
+ r_0^{-1}\OO((E-E_0)^{-\alpha})
= \delta(\eps) \, d(\vec q) + \OO(r_0^{-1+\alpha}
\eps^{-\alpha}), \qquad \alpha<1\, .
\ee
Thus in the $r_0\to\infty$ limit the correction terms vanish, and we
conclude that the $CFT_1$ living on the boundary contains only zero
energy states. $d(\vec q)$ denotes the number of such
zero energy states
and is related to the $AdS_2$ partition
function via eq.\refb{et36}. Thus the $CFT_1$
relevant for $AdS_2/CFT_1$ correspondence seems to be quite
different from the ones studied in \cite{fub1,fub2}. Examples of
quantum systems containing only a finite number of states can be
found in \cite{forge}.

Finally we consider the case where 
in the computation of the partition function on $AdS_2$
we fix the asymptotic values of the constant mode of the gauge field
$A^{(i)}_\theta$ to $ie_i$ (see eq.\refb{et8})
and allow the electric fields to fluctuate.
As already mentioned, such a partition function is not sensible
in a generic situation, but may exist in special cases.
In 
the $w$ coordinate system the field configuration
\refb{et15} near the boundary is independent of $r_0$ up to corrections
of order $r_0^{-2}$, except for the $ie_i / r_0$ term in $A_w^{(i)}$.
Let us denote
by $H$ the generator of $w$ translation in the boundary theory in the
$r_0\to \infty$ limit.
Then for finite $r_0$ the generator of $w$ translation
will be given by $H$ plus  correction
terms suppressed by inverse powers of $r_0$. Of these the $r_0^{-1}$
term will come solely from the 
$iQ_i A_w^{(i)}$
term that appears in the Hamiltonian. 
In particular the $-ie_i / r_0$ term in $A_w^{(i)}$
gives an additive term $e_i Q_i r_0^{-1}$ to the $w$ translation
generator where
$Q_i$ is the charge conjugate to $A_\mu^{(i)}$ in the boundary CFT.
Thus the partition function of the boundary CFT is given by
\be \label{et21}
Z_{CFT_1} = Tr\left(e^{-2\pi r_0 (H + e_i Q_i r_0^{-1}
+ \OO(r_0^{-2}))}\right)=
Tr\left(e^{-2\pi r_0 H - 2\pi e_i Q_i + \OO(r_0^{-1})}\right)\, .
\ee
In this case
the dual $CFT_1$
should in general contain states carrying different charges.
For large $r_0$ the partition function \refb{et21} for such a $CFT$
should reduce to
\be \label{et22}
Z_{CFT_1} = e^{-2\pi E_0 r_0} \, \sum_{\vec q} d(\vec q) \, e^{-2\pi 
\vec e \cdot \vec q}\, ,
\ee
where $E_0$ is ground state energy and $d(\vec q)$ is the number of
ground states with $Q_i$ eigenvalue $q_i$.
Equating $Z_{CFT_1}$ given in \refb{et22} to $Z_{AdS_2}$ given
in \refb{et19} we get
\be \label{et23}
E_0 = -v \, 
\LL^{(2)}_{eff}-{K'\over 2\pi}, \qquad
Z_{AdS_2}^{finite} = 
\sum_{\vec q} d(\vec q) \, e^{-2\pi 
\vec e \cdot \vec q}\, .
\ee

\sectiono{$AdS_2/CFT_1$ Correspondence: The Microscopic
View} \label{s4}

In the last section we have used the partition function of quanum
gravity on $AdS_2$
to define the partition function of a $CFT_1$ living on the boundary
of $AdS_2$. There is however another aspect of $AdS/CFT$ 
correspondence that allows us to give an independent description of
the boundary CFT: it is the low energy dynamics of the system of branes
which produce the particular $AdS$ space as its near horizon 
geometry\cite{9711200}.
Since the $AdS_2$ geometry under consideration arises as the near
horizon geometry of an extremal black hole, we expect that the 
corresponding $CFT_1$ will describe the low energy dynamics of the
brane system which describes this black hole. However for all known 
cases the spectrum of the brane system that produces the black hole
has a gap that separates the ground state from the first excited state
in any given charge sector. Thus one would expect that in the low
energy limit all the excited states will disappear and
the $CFT_1$ will
capture information about only the ground states of the brane system.
After suitable shift in the energy that sets the ground state energy in
each charge sector to zero, we find that $CFT_1$ contains only zero
energy states, in agreement with the results of the last section. 
Furthermore if $d_{micro}(\vec q)$
denotes
the number of such ground states of the brane system
in a given charge sector, then identifying $CFT_1$ with the low
energy dynamics of the brane system leads to
the identification
\be \label{et39}
d(\vec q) = d_{micro}(\vec q)\, .
\ee 
This gives from \refb{ettex}
\be \label{ett1a}
d_{micro}(\vec q) = \left\langle \exp[-
i  q_i\ointop d\theta \, A^{(i)}_\theta]
\right\rangle^{finite}_{AdS_2}\, .
\ee
On the other hand if we define the $AdS_2$ partition function by
fixing the constant modes of the gauge fields
at the boundary,
then we get from \refb{et23},
\be \label{et40}
Z_{AdS_2}^{finite} = 
\sum_{\vec q} d_{micro}(\vec q) \, e^{-2\pi 
\vec e\cdot \vec q}\, .
\ee
Eqs.\refb{ett1a} and \refb{et40}  reproduce the results 
\refb{ealt} and \refb{egen1}
quoted in the introduction.

\sectiono{The Classical Limit} \label{s4.8}

We shall now take the classical limit of \refb{ealt} and show that
the result reduces to the statement of equality between Wald
entropy and the statistical entropy. 
For this we note that in the classical limit we can replace the right
hand side of \refb{ealt} by the product of the classical $AdS_2$
partition function \refb{et17} and 
$\exp[-i  q_i\ointop d\theta \, A^{(i)}_\theta]$ with 
$A^{(i)}_\theta$
given by its classical value \refb{et8}. 
After removing terms in the exponential linear in
$r_0$, eq.\refb{ealt} reduces to
\be \label{ered1}
d_{micro}(q) = e^{2\pi \vec q\cdot \vec e - 2\pi v \LL^{(2)}}\, .
\ee
Taking the logarithm on the two sides we get 
\be \label{et43pre}
 \ln d_{micro}(\vec q) = 2\pi \left( \vec e \cdot \vec q
 -v \, \LL^{(2)} 
  \right)\, .
 \ee
The left hand side gives the statistical entropy whereas the
right hand side gives the classical
Wald entropy according to \refb{et3}, \refb{et5}.

Next we consider the classical limit of \refb{egen1}.
First of all in this limit 
the left hand side is
given by \refb{et18}.
On the other hand we expect the right hand side to be sharply peaked
as a function of $\vec q$ and hence the leading contribution will be
given by the value of the summand at its extremum. This leads to
the relation
\be \label{et41}
 -2\pi v \, \LL^{(2)} 
  = \ln d_{micro}(\vec q) - 2\pi \vec e \cdot \vec q \, ,
 \ee
 at
 \be \label{et42}
 {\p\, \ln d_{micro}(\vec q) /\p q_i = 2\pi e_i}\, .
 \ee
 This is equivalent to
 \be \label{et43}
 \ln d_{micro}(\vec q) = 2\pi \left( \vec e \cdot \vec q
 -v \, \LL^{(2)} 
  \right)\, ,
 \ee
 at
 \be \label{et44}
 q_i = {\p\over \p e_i} \left(v \LL^{(2)}\right)\, .
 \ee
 Comparing this with \refb{et3}-\refb{et5} we get
 \be \label{et45}
 S_{BH} = \ln d_{micro}(\vec q)\, .
 \ee

 Before concluding this section we would like to discuss the
 precise meaning of the classical limit. In any theory this can be
 done by multiplying $\LL^{(2)}$ by a scale factor $\lambda$
 and then taking $\lambda$ to be large. In string theory this can be
 achieved by a redefinition of the dilaton field $\phi$ 
 and the RR fields.
 In particular we need to scale $e^{-2\phi}$ to $\lambda e^{-2\phi}$,
 and any $RR$ field $\psi_{RR}$ to $\lambda^{1/2}\psi_{RR}$,
 leaving the 
 NSNS sector fields (other than the dilaton)
 unchanged. For a black hole in (3+1) dimensions
 this induces the
 transformation
 \be \label{exk1}
 p^{NSNS}\to p^{NSNS}, \quad e^{NSNS}\to e^{NSNS},
 \quad p^{RR}\to \lambda^{1/2} p^{RR}, \quad 
 e^{RR}\to \lambda^{1/2} e^{RR}\, ,
 \ee
 where $p$ denotes the magnetic charges (which are hidden arguments
 of $Z_{AdS_2}$) and $e$ denotes the electric fields. Under this scaling
 $q_i$ computed through \refb{et44} scales as
 \be \label{exk2}
 q^{NSNS}\to\lambda \, 
 q^{NSNS}, \quad q^{RR}\to \lambda^{1/2}
 q^{RR}\, .
 \ee
 Thus in \refb{egen1} the classical limit will correspond to scaling the
 magnetic charges and the electric fields in the argument of 
 $Z^{finite}_{AdS_2}$ as in \refb{exk1} and then taking the
 large $\lambda$ limit. On the other hand 
 in \refb{ealt} we need to scale 
 $\vec p$ and $\vec q$ in the argument of
 $d_{micro}$ as in \refb{exk1}, \refb{exk2}.

\sectiono{OSV Conjecture and $AdS_2/CFT_1$ Correspondence}
\label{s5.5}

\def\FF{F}

$AdS_2/CFT_1$ correspondence suggests a possible route
to deriving the OSV conjecture\cite{0405146} that relates the
partition function for BPS states in $\NN=2$ supersymmetric string
theories to the topological 
string partition function.\footnote{Alternate approaches to 
analyzing the black hole partition function and deriving the
OSV conjecture using 
AdS/CFT correspondence can be found in
\cite{0607138,0605279,0608059,0704.0955,0704.1405}.
The advantage of 
our approach lies in the fact that
since we apply $AdS/CFT$ correspondence on the near
horizon geometry with the $AdS_2$ factor, 
the chemical potentials dual to the charges
are directly related to the near horizon electric field, and hence,
via the attractor mechanism, to other near horizon field
configuration. 
Furthermore the path integral needs to be performed only
over the near horizon geometry where we have enhanced 
supersymmetry and hence stronger non-renormalization
properties. Ref.\cite{0608021} suggests an approach based purely
on the analysis of the partition function in the near horizon geometry;
however their approach to dealing with the divergence from infinite
volume of $AdS_2$ is quite different from ours.}
For this we note that the right hand side
of \refb{egen1} is the black hole partition function
defined in \cite{0405146}.
 Thus if we can
evaluate $Z_{AdS_2}^{finite}$ we shall compute the
black hole partition function. Now the effective action of
$\NN=2$ supergravity theories in four dimensions contains a special
class of terms, known as F-terms. Since these terms are local and do
not require any infrared regulator for their definition, we
expect that the structure of these terms should not 
depend on the background in which they are evaluated, and part of
$\LL^{(2)}_{eff}$ appearing in \refb{et20} 
should be given just by these
F-terms evaluated in the attractor geometry.
The information about the
`F-type terms' can be encoded in a function
$\FF(\{ X^I\}, \wh A)$ -- known as the generalized prepotential --
of a set of complex variables $X^I$
which are in one to one correspondence with the gauge fields and
an auxiliary complex variable $\wh A$ related to the square of
the graviphoton field strength\cite{9602060,9603191}. 
$\FF$ is a homogeneous function of degree two in its
arguments:
\be \label{efcon}
\FF(\{\lambda X^I\}, \lambda^2 \wh A) = \lambda^2
\FF(\{ X^I\}, \wh A)\, .
\ee
For a given choice of electric field one finds that
the 
extremum of the effective Lagrangian density computed with the
$F$-term effective action occurs at the 
attractor point
where\cite{9508072,9602111,9602136,
9711053,9801081,9812082,0007195}
\be \label{eatt}
\wh A=-4w^2, \quad 
4(\bar w^{-1} \bar X^I + w^{-1}  X^I) = e^I, \quad
4(\bar w^{-1} \bar X^I - w^{-1}  X^I) = - i p^I\, .
\ee
Here $w$ is an arbitrary complex parameter and $p^I$ are the
magnetic charges carried by the black hole. These magnetic
charges have not
appeared explicitly in our discussion so far because from the
point of view of the near horizon geometry they represent
fluxes through compact two cycles and
appear as parameters labelling
the two (or three) 
dimensional field theory describing the near horizon
dynamics.
The value of the effective Lagrangian density at the
extremum \refb{eatt} is given by\cite{0603149}
\be \label{eatt2}
v\LL^{(2)}_{eff}= 16 \, i \, (w^{-2}\FF - \bar w^{-2}\bar \FF)\, .
\ee
Note that \refb{eatt} determines $X^I$ in terms of the
unknown parameter $w$. However due to the scaling symmetry
\refb{efcon}, $v\LL^{(2)}_{eff}$
given in \refb{eatt2} is independent of $w$.
Using
this scaling symmetry we can choose
\be \label{eatt3}
w = - 8 \, i\, ,
\ee
and rewrite \refb{eatt}, \refb{eatt2} as
\be \label{eatt4}
\wh A=256, \quad X^I = - i (e^I + i p^I)\, ,
\ee
\be \label{eatt5}
v\LL^{(2)}_{eff}= -{i\over 4} 
(\FF(\{X^I\}, 256) - \overline{\FF(\{X^I\}, 256)})
\, .
\ee
Thus the contribution to $Z_{AdS_2}$ from the F-terms is given by
\be \label{ezexp}
Z^F_{AdS_2} = e^{-\pi \, Im \, \FF(\{p^I - i e^I\}, 256)}
\, .
\ee
If we assume that this is the full contribution to $Z_{AdS_2}$ then
eqs.\refb{egen1}, \refb{ezexp} leads to
the original OSV conjecture\cite{0405146}.

It has however been suggested in subsequent papers that agreement
with statistical entropy requires modifying this formula by including
additional measure factors on the right hand side
of \refb{ezexp}\cite{0508174,0601108,0702146,0808.2627}. 
A careful analysis
of the path integral keeping track of the holomorphic 
anomaly\cite{9302103,9307158,9309140}
may be able to
reproduce these corrections.
Some of these corrections are in fact
necessary for restoring
the duality invariance of the final result for the 
entropy\cite{0601108,0808.2627}. 

Note that in trying to prove OSV conjecture from $AdS_2/CFT_1$
correspondence we must work with the partition function 
$Z_{AdS_2}$. As mentioned earlier this may not always be
defined as it requires integrating over fluctuations of the
non-normalizable modes representing asymptotic electric fields,
but $\NN=2$ supersymmetry may  help in making
$Z_{AdS_2}$ well defined.

\sectiono{Conclusion, Caveats  and Open Questions} \label{s6}

The concrete achievements of the analysis given above can be
summarized as follows.

\begin{enumerate}

\item We have been able to give 
a proper definition
of the partition function of a black hole in near horizon
attractor geometry of an extremal black hole. Typically such
partition functions suffer from infrared divergence, 
but we have described
an unambiguous procedure for extracting its finite 
part.\footnote{Typically such partition function also suffers
from ultraviolet divergence but we expect string theory to
cure them. If we have sufficient amount of supersymmetry
then these ultraviolet divergences can also cancel due to
supersymmetry.} 

\item 
The usual rules of
$AdS_2/CFT_1$ correspondence relates the partition
function of quantum gravity on $AdS_2$ to
the partition function of a $CFT_1$ living on the boundary of
$AdS_2$.
While we have not given an explicit Lagrangian 
description of this CFT, the usual rules of $AdS/CFT$
correspondence allows us to identifiy it as the infrared limit
of the quantum mechanics associated with the brane system that
describes the black hole. Since in any given charge sector the
spectrum has a gap that separates the ground state from the excited
states, the Hilbert space of
$CFT_1$ consists of only the ground states of this quantum
mechanical system. 

\item
Whether a $CFT_1$ dual to a given attractor
geometry contains only the ground states of a given charge associated
with the attractor geometry or ground states for all charges 
depends on whether in the definition of the $AdS_2$ 
partition function
we have fixed the asymptotic values of the electric fields 
 or the asymptotic values of the constant modes of the gauge
fields. The former is more
natural since electric field mode is the dominant (non-normalizable)
mode of the gauge field near the boundary. In this case the 
dual $CFT_1$ would contain states of a fixed charge, and we are led
to \refb{ealt}. On the other hand if we fix the asymptotic values of the
constant modes of the gauge fields then in the definition of the dual
$CFT_1$ we must include sum over states carrying different charges.
This leads to \refb{egen1}. However since this involves integrating
over non-normalizable modes of $AdS_2$, such a partition function
may not always be well defined. Nevertheless this partition function 
is what appears
naturally in the OSV conjecture; so in theories with enhanced 
supersymmetry such partition functions may also be well-defined.

\item In the
classical limit both \refb{ealt} and \refb{egen1} reduce 
to the usual relation between
$\ln d_{micro}(\vec q)$ and the Wald entropy. 
Thus based on our analysis one can conclude that
the equality of the classical Wald entropy
and the statistical entropy of an extremal 
black hole for large charges is a direct consequence of
$AdS_2/CFT_1$ correspondence in the classical limit.
\end{enumerate}
Our analysis does not
use supersymmetry directly, although supersymmetry is
undoubtedly useful in stabilizing the extremal black holes
against quantum corrections.

There are however many caveats, many things which require
clarification and many open questions. Some of them have already
been mentioned earlier, but for the benefit of the reader we shall now
summarize these issues.

\begin{enumerate}

\item Relation  \refb{ealt} requires
us to define the partition function on $AdS_2$
for a given set of values of
electric charge $\vec q$ , while relation
\refb{egen1} requires us to be able
to define it for a given set of values
of the constant modes of the asymptotic gauge fields. In the
classical limit the latter are equal to the near horizon
electric fields $\vec e$. 
Although generically $\vec e$ is fixed by
$\vec q$ and vice versa, in many examples in classical string theory
treating $\vec e$ as independent variables
can be problematic. Consider for example an
heterotic
string theory  compactification. In this case after extremizing
the near horizon classical
Lagrangian density with respect to all the fields except the dilaton
$\phi$, $v\LL^{(2)}$ takes the form $e^{-2\phi} f(\vec e)$ for some
function $f(\vec e)$. Extremizaion with respect to the dilaton
now gives $f(\vec e)=0$, \i.e.\ it gives a constraint among the $e_i$'s
instead of fixing the dilaton. This makes defining 
$Z_{AdS_2}(\vec e)$ problematic -- at least in the classical limit --
since for this we need to extremize $v\LL^{(2)}$ with respect to
all the near horizon parameters other than the electric fields and express
the result as a function of $\vec e$.
If instead we fix $\vec q$, then we do not encounter such problems
since $\phi$ as well as the $e_i$'s can be computed in terms of $\vec q$
via the extremization equation.

Due to this \refb{egen1} seems to be ill defined for such theories,
whereas \refb{ealt} does not have such difficulties. This however
is not a serious problem as one can use a dual description, exchanging
the roles of some of the electric and magnetic charges, where both
prescriptions can be made well defined.

\item Eq.\refb{egen1} is close in spirit to the OSV conjecture, but as a
result it also shares some of its problems. One of these is that in
many supersymmetric string theories the sum over $\vec q$ appearing
in \refb{egen1} is not convergent since the function 
$\ln d_{micro}(\vec q)$, when Taylor
expanded around some given $\vec q_0$,
does not have a negative definite quadratic term. As in the case of
OSV proposal, we can define this by formally continuing the sum
over $\vec q$ to complex values\cite{0508174}. While this allows
us to carry out an asymptotic expansion, it is not clear
if this procedure can be made exact. In contrast \refb{ealt}, not having
any sum over $\vec q$, does not suffer from such problems.

\item For an extremal black hole the ground state degeneracy
$d_{micro}(\vec q)$ is not expected to change continuously
as we vary the asymptotic moduli. However discrete jumps are
possible and they occur as we cross walls of marginal stability.
Thus in order to make \refb{egen1} or \refb{ealt} well-defined
we need to  decide on a precise prescription for
$d_{micro}(\vec q)$ that appears in these equations.  In
(3+1) dimensional flat space-time the natural
choice for this seems to be the value $d_{micro}(\vec q)$ when
the asymptotic moduli are set equal to their attractor values. In
this case on the black hole side the contribution to the entropy
is known to come solely from single centered black holes with
a single $AdS_2$ factor in their near horizon 
geometry. 
On the
other hand as we cross a wall of marginal stability new
multi-centered solutions carrying the same total charge and mass
appear, and $d_{micro}(\vec q)$ in such a background should count
the total degeneracies of single and multi-centered black 
holes\cite{0005049,0010222,0101135,0206072,
0304094,0702141,0702146,0702150,0705.3874,0706.2363}. 
Since
our analysis has been based on the partition function of a single
$AdS_2$, it is only natural that $Z_{AdS_2}$ will only contain
information about single centered black holes.

\item Although we have followed the spirit of $AdS_{d+1}/CFT_d$ 
correspondence, the actual implementation of this for $d=1$
is somewhat different from that for higher $d$. For any $d$ the
euclidean version of $AdS/CFT$ correspondence relates the
partition function on euclidean global $AdS_{d+1}$ to partition 
function of $CFT_d$ living on the boundary 
$S^d$\cite{9802109,9802150}. However only in 
the special case of $d=1$ the CFT
partition function, being on $S^1$, may be represented as a trace over
the Hilbert space of the CFT. 
This is what we have exploited in our analysis. The time coordinate
labelling the coordinate of the boundary circle can be identified
(up to a scaling) to the Schwarzschild time of the black hole.

\item
For $d\ge 2$ one follows a different route for relating
the CFT spectrum to  quantum gravity on $AdS_{d+1}$. One
uses the exponential map to represent global $AdS_{d+1}$
as $B^d\times \RRR$  and its boundary as 
$S^{d-1}\times \RRR$. The
parameter labelling $\RRR$ is the global $AdS_{d+1}$ time, and this
allows us to identify the Hilbert space of states of $CFT_d$ 
on
$S^{d-1}\times \RRR$ to the Hilbert space of states
of quantum gravity on $AdS_{d+1}$, with the
Hamiltonian of the $CFT_d$ getting mapped 
to the generator of global time translation in $AdS_{d+1}$.
For $d=1$ however $S^{d-1}$ corresponds to a pair of disconnected
points and hence  we have two copies of $CFT_d$, one at each
boundary. It has been suggested that the
extremal black hole entropy can be related to the entanglement
entropy of this pair of $CFT$'s\cite{0710.2956}. 
Is there a relation between this proposal and the one we have
presented here? To this end note that according to our proposal
in the sector with charge $\vec q$ the $CFT_1$ has precisely 
$d(\vec q)$ states, all with zero energy. Now if we assume that
the ground state of string theory on $AdS_2$, represented
as $B^1\times \RRR$, corresponds to a
state in the boundary $CFT_1$ where
the two copies of the $CFT_1$ living on the two boundaries are
maximally entangled, then the entanglement entropy will be given
by $\ln d(\vec q)$, in agreement with the statistical entropy
of one of the boundary $CFT_1$'s.
Indeed for extremal BTZ black holes precisely 
such a route to relating the entanglement entropy and statistical
entropy was suggested in \cite{0710.2956}.
It will be interesting to explore this relationship directly in the bulk
theory. In the classical limit both proposals reduce to the Wald entropy,
but in the quantum theory they may differ.

\end{enumerate}

\medskip

\noindent {\bf Acknowledgement:} I would like to thank 
Atish Dabholkar, Justin David,
Rajesh Gopakumar, Dileep Jatkar, Juan Maldacena,
Shiraz Minwalla, Eliezer Rabinovici and
Sandip Trivedi for useful discussions and 
Rajesh Gupta for useful discussion and collaboration
in \cite{0806.0053}. I would also
like to thank the organisers of Strings 2008 conference
for providing
a stimulating environment where some of these 
results were presented.



\end{document}